\documentclass[computers,article,submit,moreauthors, pdftex,10pt,a4paper]{mdpi} 
\usepackage{graphicx}
\usepackage{multirow}
\usepackage{amsmath,subfig}

\firstpage{1} 
\articlenumber{x}
\doinum{10.3390/------}
\pubvolume{xx}
\pubyear{2016}
\copyrightyear{2016}
\externaleditor{Academic Editor: name}
\history{Received: date; Accepted: date; Published: date}
\pdfoutput=1

 \theoremstyle{mdpi}
 \newcounter{thm}
 \newcounter{ex}
 \newcounter{re}
 \theoremstyle{mdpidefinition}

\Title{Visual Motion Onset Brain--computer Interface}

\Author{Jair Pereira Junior $^{1,2,\dagger}$, Caio Teixeira $^{1,3,\dagger}$ and Tomasz M. Rutkowski $^{1,\ddagger,}$*}
\AuthorNames{Jair Pereira Junior, Caio Teixeira and Tomasz M. Rutkowski}

\address{%
$^{1}$ \quad University of Tsukuba, Tsukuba, Japan\\
$^{2}$ \quad Federal University of ABC (UFABC), Santo Andre, Sao Paulo, Brazil\\
$^{3}$ \quad University of Campinas (UNICAMP), Campinas, Sao Paulo, Brazil}

\corres{Correspondence: tomek@bci-lab.info}

\firstnote{These authors contributed equally to this work.}
\secondnote{Current address: The University of Tokyo, Tokyo, Japan} 

\abstract{The paper presents a study of two novel visual motion onset stimulus-based brain--computer interfaces (vmoBCI). Two settings are compared with afferent and efferent to a computer screen center motion patterns. Online vmoBCI experiments are conducted in an oddball event--related potential (ERP) paradigm allowing for ``aha--responses'' decoding in EEG brainwaves.
A subsequent stepwise linear discriminant analysis classification (swLDA) classification accuracy comparison is discussed based on two inter--stimulus--interval (ISI) settings of $700$ and $150$~ms in two online vmoBCI applications with six and eight command settings.
A research hypothesis of classification accuracy non--significant differences with various ISIs is confirmed based on the two settings of $700$~ms and $150$~ms, as well as with various numbers of ERP response averaging scenarios. The efferent in respect to display center visual motion patterns allowed for a faster interfacing and thus they are recommended as more suitable for the no--eye--movements requiring visual BCIs.}


\begin{document}

\section{Introduction}

A brain--computer interface (BCI) is a neurotechnology that utilizes only central nervous system (CNS) signals (brainwaves) of a user to create a new communication channel with others or to control external devices without depending on any muscle activity~\cite{bciBOOKwolpaw}. 
The BCI technology has provided a support already to patients' life improvement who suffer from severe paralysis due to diseases like an amyotrophic lateral sclerosis (ALS)~\cite{bciBOOKwolpaw}. 
The successful alternative options have been developed recently to utilize spatial
auditory~\cite{MoonJeongBCImeeting2013,iwpash2009tomek} or tactile (somatosensory) modes~\cite{tomekJNM2015,HiromuBCImeeting2013}.
Meanwhile, the visual BCI seems to offer still the superior communication options in comparison with the above modalities~\cite{martinez2007fully,bciBOOKwolpaw,auditoryBCIp300PREDICTION2013,sakurada2014use,daikiBIH2015,daikiEMBC2015,DaikiBCImeeting2016}, yet they require intentionally controlled eye--movements, which some of locked--in syndrome (LIS) users cannot maintain~\cite{lis139review1986,tactileAUDIOvisualBCIcompare2013}.

\begin{figure}%[th]%
    \centering
   	\subfloat[A screenshot from a video available online (\url{https://youtu.be/JbdazbFntek}) presenting the afferent visual motion onset BCI paradigms with stimuli trajectories towards the screen center]{
        \includegraphics[width=12cm]{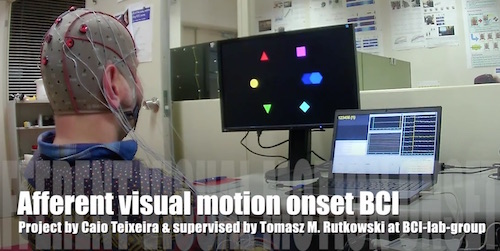}\label{fig:affVMOBCI}
    }
    \\ 
    \subfloat[A screenshot from a video available online (\url{https://youtu.be/UaxIQdOctcQ}) presenting the efferent visual motion onset BCI paradigms with stimuli trajectories towards the screen center]{
    	\includegraphics[width=12cm]{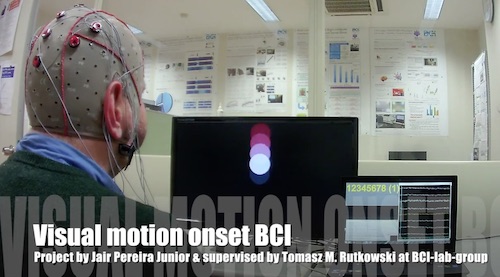}\label{fig:effVMOBCI}
	}
    \caption{The user wearing EEG cap seats in front of a computer display with the visual motion onset BCI paradigms. {\bf(a)} An afferent visual motion case with six stimuli. {\bf(b)} An efferent visual motion case with eight patterns. In the both photographs the EEG brainwave signals are presented on smaller computer displays with numbers representing the target movements to be recognized by the user. Published with permission of the depicted user.}
    \label{fig:f1}
\end{figure}

We present results of a novel study utilizing two strategies of the visual motion onset patterns~\cite{kremlavcek2004effect,guo2008brain} for BCI purposes with stimuli presented as afferent or efferent to a centre of the field of vision, the fovea.
The vmoBCI patterns used in the presented experiments are shown in Figures~\ref{fig:affVMOBCI}~and~\ref{fig:effVMOBCI} for afferent and efferent movements, respectively. 
A goal of this study is to compare and test a BCI performance (a classification accuracy) of the novel vmoBCI paradigm in function of two inter--stimulus interval (ISI) settings. Namely the ISI equal to $700$~ms (a very easy case) and $150$~ms (a harder case due to fast repetition of the presented visual movement patterns and larger overlap of the brainwave ERPs) are evaluated in the paper.

Results obtained with ten healthy users in online BCI experiments with stepwise linear discriminant analysis (swLDA) classifier~\cite{krusienski2006} and two different averaging settings are analyzed and discussed.
From now on the paper is organized as follows. In the next section we present results.
Discussion, materials and methods used in order to capture, process and classify the brainwave response in application to the proposed vmoBCI follow. Conclusions summarize the paper.
%

%%%%%%%%%%%%%%%%%%%%%%%%%%%%%%%%%%%%%%%%%%
\section{Results}

\begin{figure}%[ht]%
    \centering
   	\subfloat[$ISI=700$~ms.]{
        \includegraphics[height=10cm]{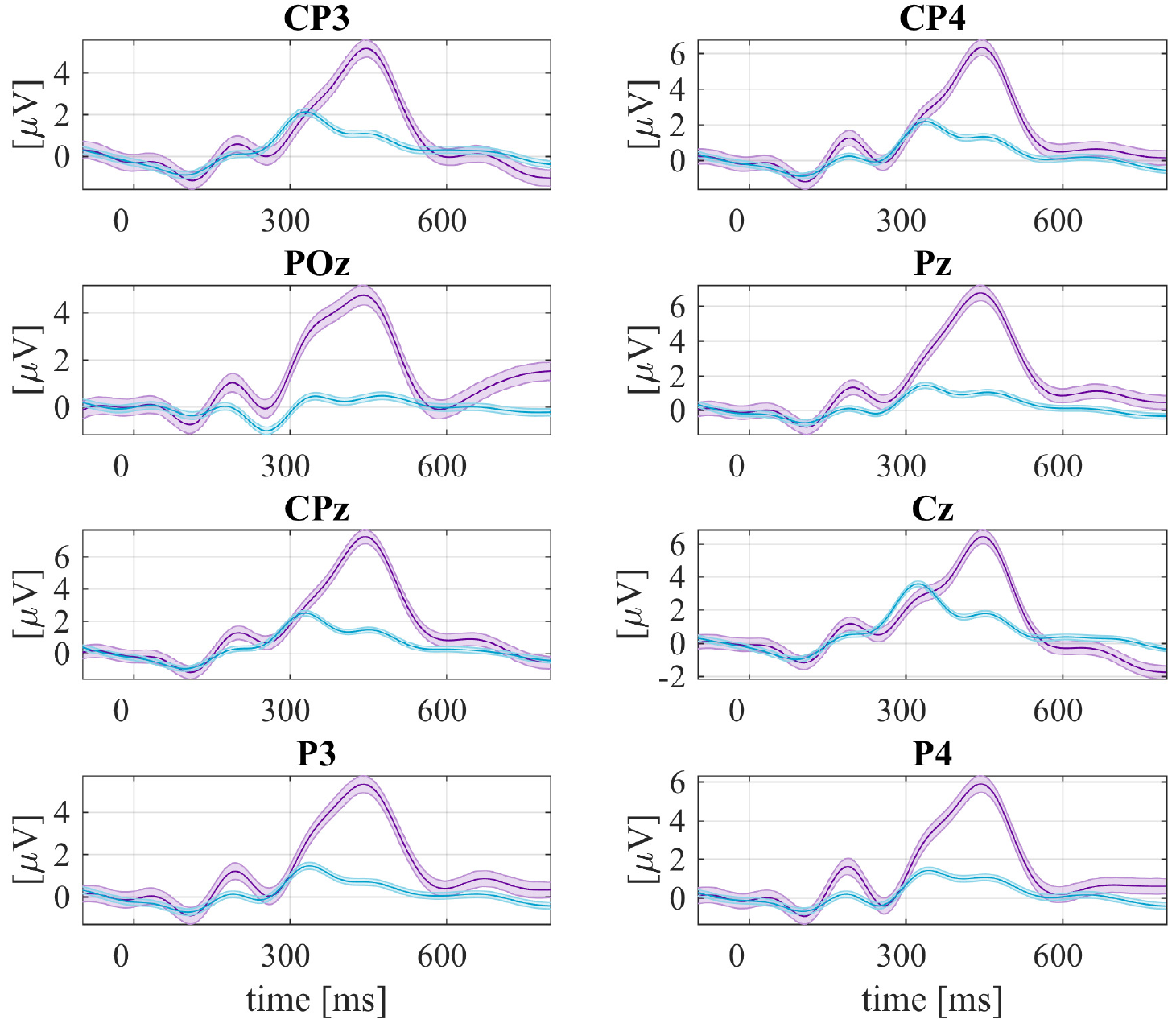}\label{fig:erpAFF700}
    }  
    \\ 
    \subfloat[$ISI=150$~ms.]{
    	\includegraphics[height=10cm]{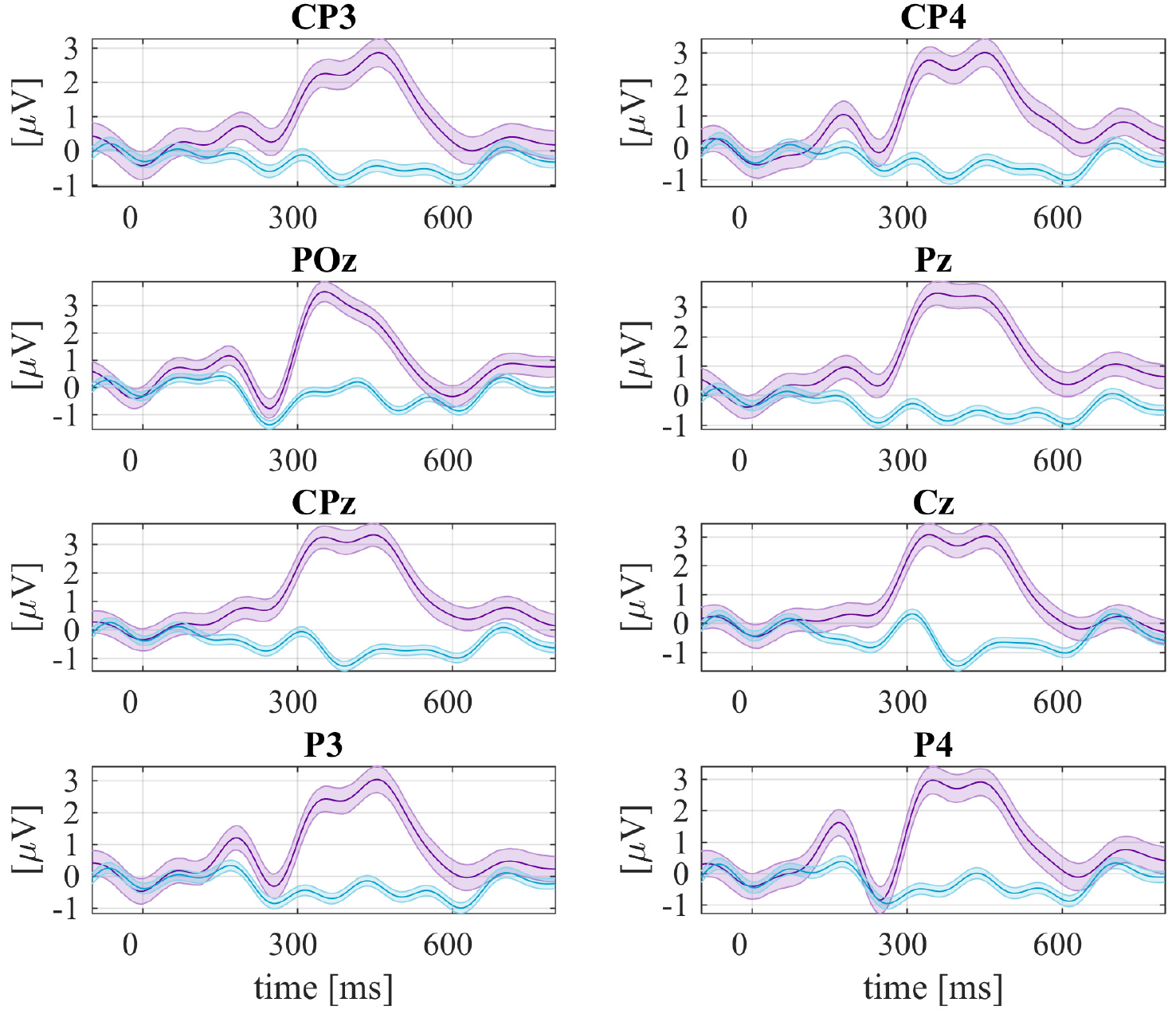}\label{fig:erpAFF150}
	}
    \caption{Grand mean averaged brainwave ERP responses in afferent vmoBCI. The green traces depict the attended targets with P300 responses and blue the non--targets. The intervals surrounding the mean traces depict standard errors.}\label{fig:erpAFF}
\end{figure}

\begin{figure}%[t]%
    \centering
   	\subfloat[$ISI=700$~ms.]{
        \includegraphics[height=10cm]{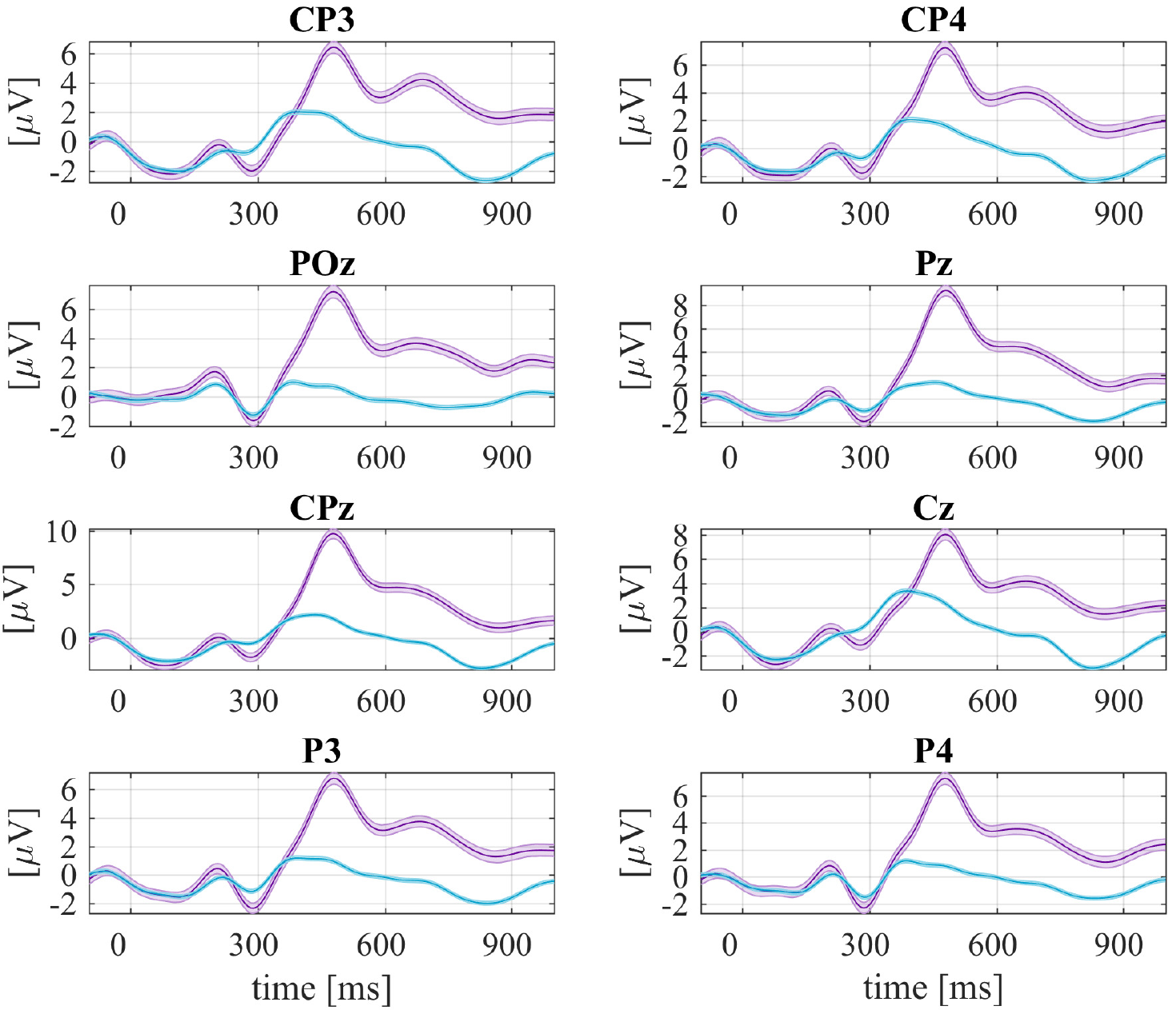}\label{fig:erpEFF700}
    }  
    \\ 
    \subfloat[$ISI=150$~ms.]{
    	\includegraphics[height=10cm]{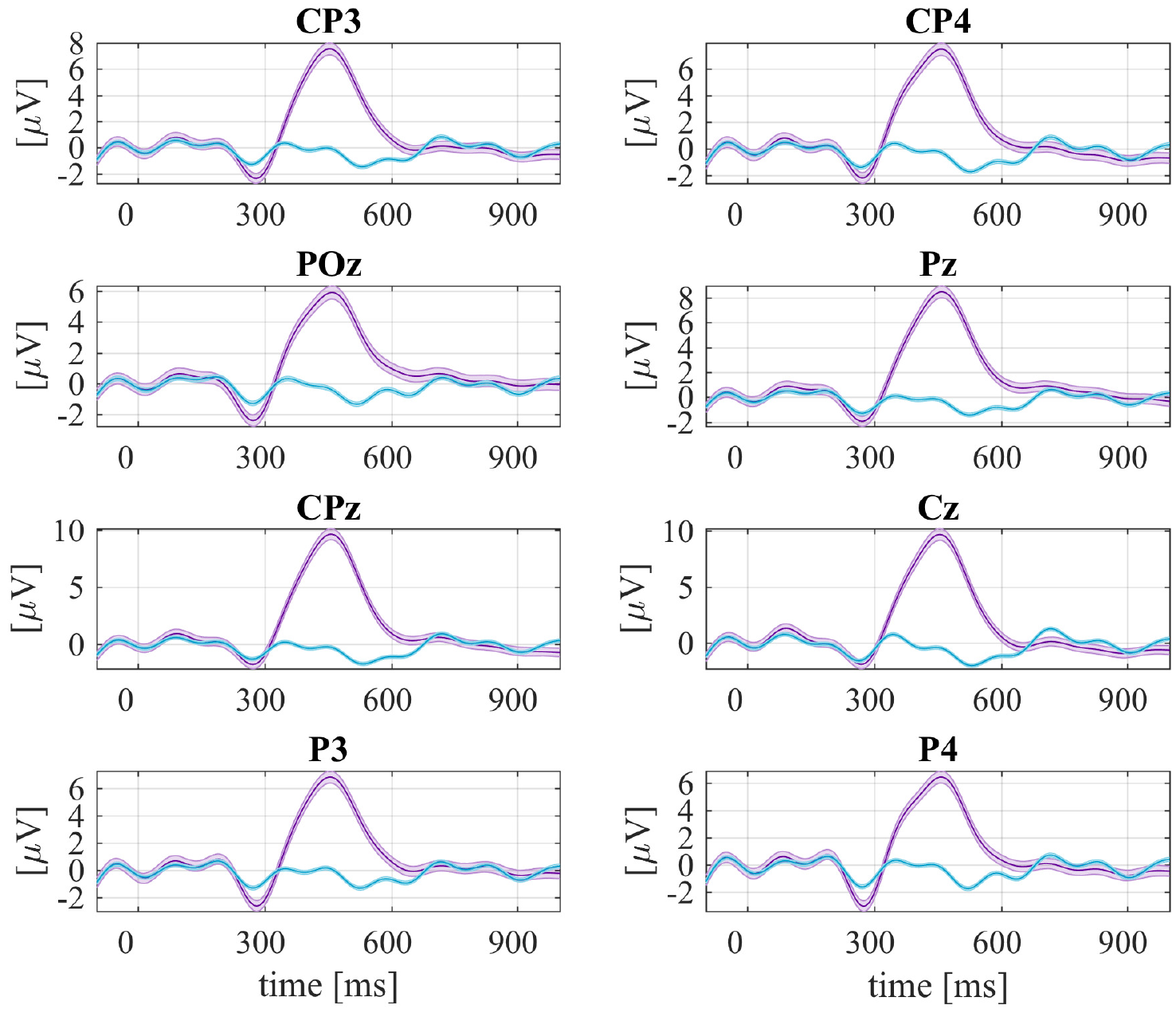}\label{fig:erpEFF150}
	}
    \caption{Grand mean averaged brainwave ERP responses in efferent vmoBCI. The green traces depict the attended targets with P300 responses and blue the non--targets. The intervals surrounding the mean traces depict standard errors.}\label{fig:erpEFF}
\end{figure}

The results of brainwave ERPs have been summarized as simple grand mean averaged time series with standard error bars in Figures~\ref{fig:erpAFF700},~\ref{fig:erpAFF150},~\ref{fig:erpEFF700},~and~\ref{fig:erpEFF150}. Those results have clearly shown a very small impact of the two tested $ISI$s of $700$~ms and $150$~ms, as well as the afferent and efferent visual motion settings summarized in Figures~\ref{fig:erpAFF}~and~\ref{fig:erpEFF}, respectively.

\begin{figure}%[t]%
    \centering
   	\subfloat[$ISI=700$~ms.]{
        \includegraphics[height=10cm]{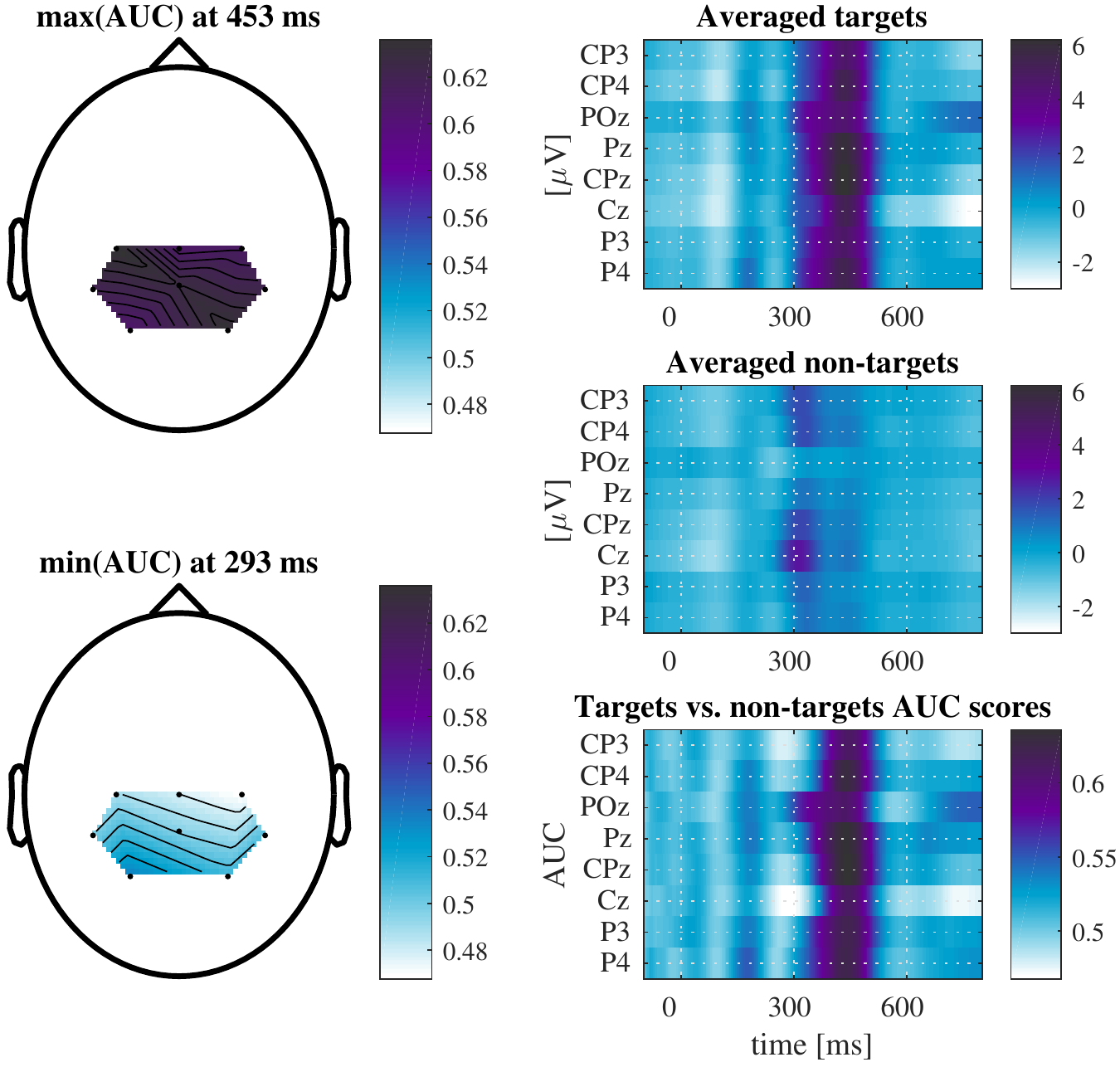}\label{fig:aucAFF700}
    }  
    \\ 
    \subfloat[$ISI=150$~ms.]{
    	\includegraphics[height=10cm]{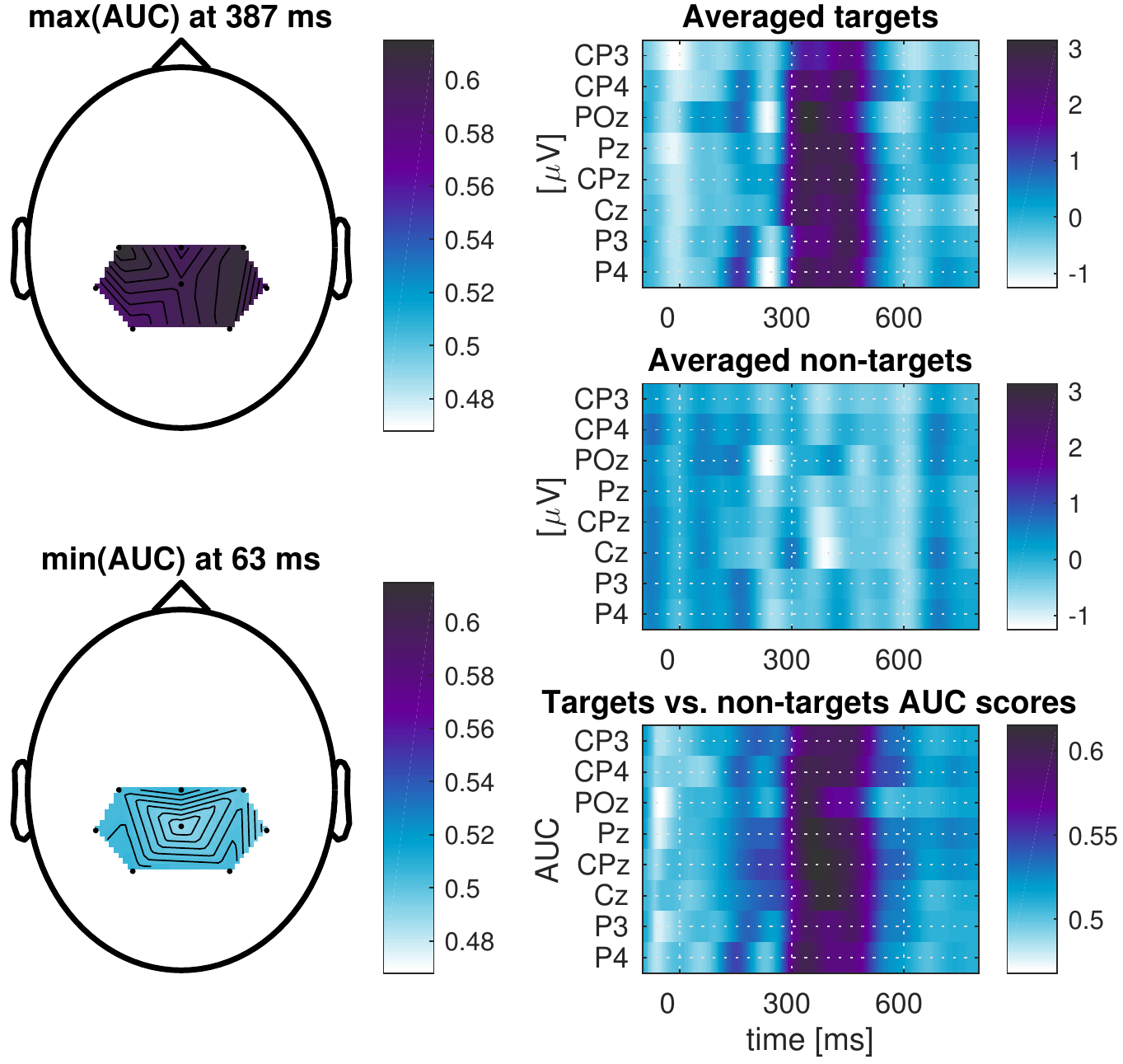}\label{fig:aucAFF150}
	}
    \caption{AUC analysis of ERP responses in afferent vmoBCI. The left column head topographic plots depict the latencies with the best (maximum) and worse (minimum) of AUC scores. The right column summarizes the ERP responses in form of matrices of all channels and AUC traces at the bottom.}\label{fig:aucAFF}
\end{figure}

\begin{figure}%[!h]%
    \centering
   	\subfloat[$ISI=700$~ms.]{
        \includegraphics[height=10cm]{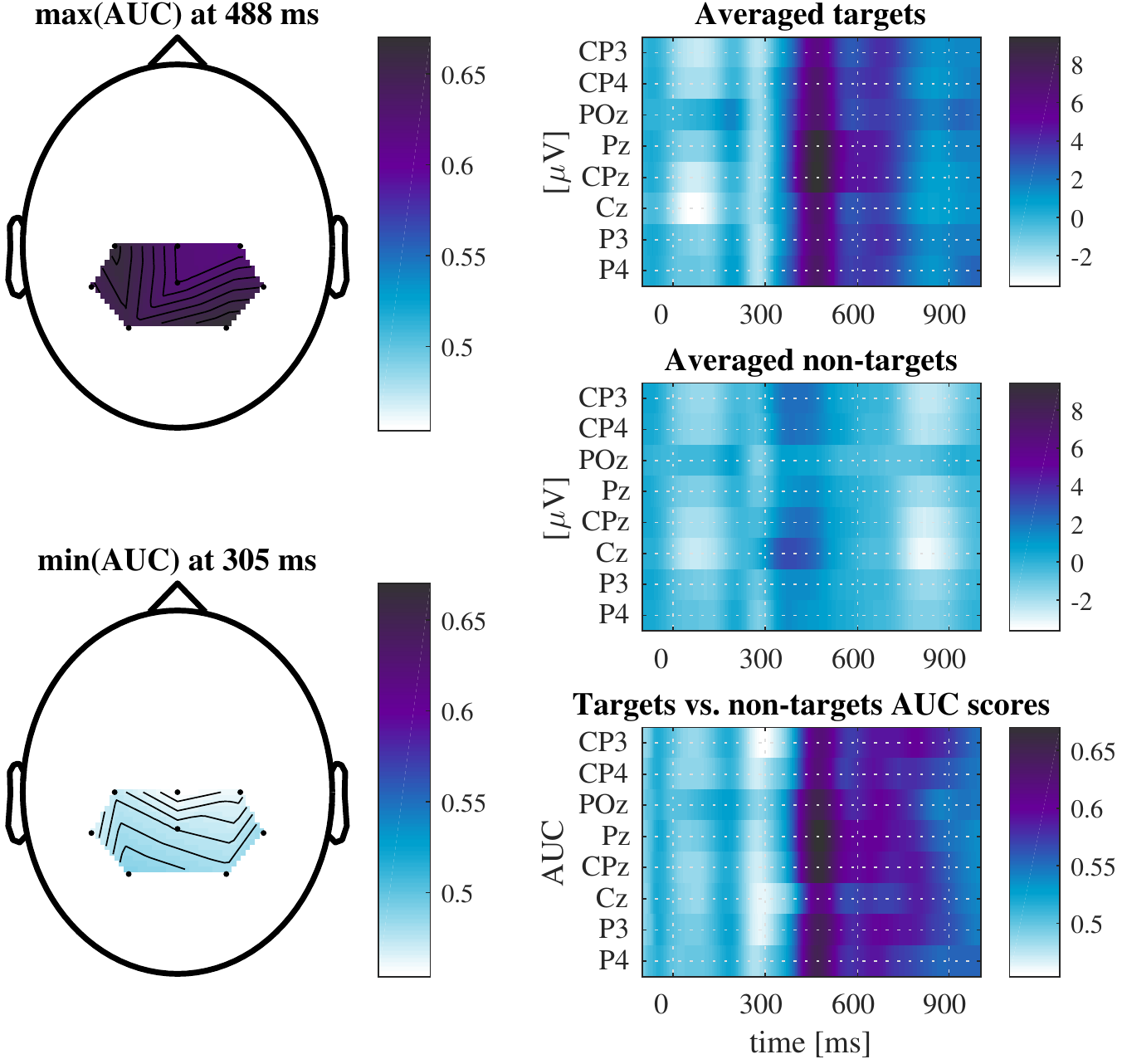}\label{fig:aucEFF700}
    }  
    \\ 
    \subfloat[$ISI=150$~ms.]{
    	\includegraphics[height=10cm]{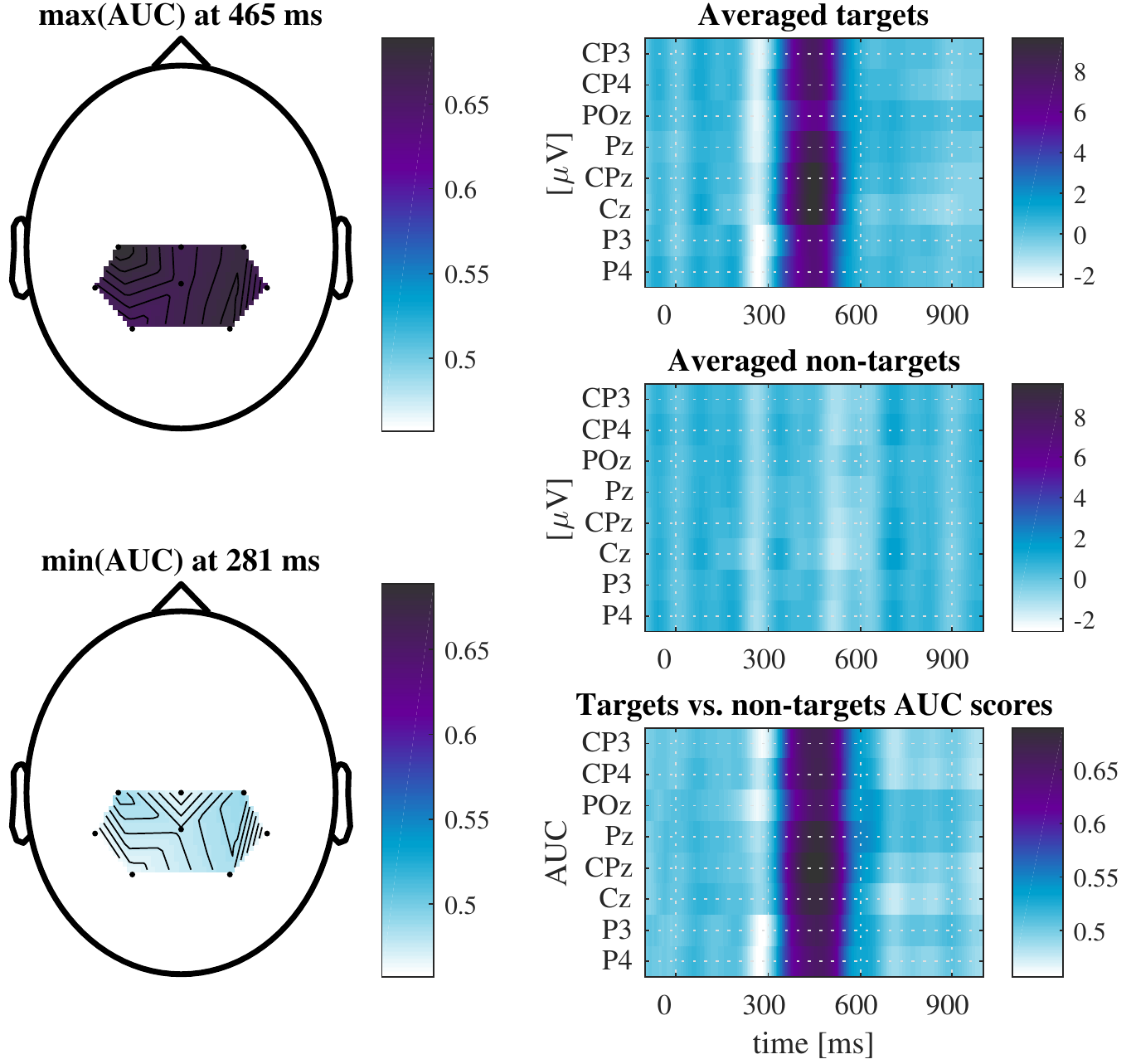}\label{fig:aucEFF150}
	}
    \caption{AUC analysis of ERP responses in efferent vmoBCI. The left column head topographic plots depict the latencies with the best (maximum) and worse (minimum) of AUC scores. The right column summarizes the ERP responses in form of matrices of all channels and AUC traces at the bottom.}\label{fig:aucEFF}
\end{figure}

More interesting results have been elucidated with an area under curve (AUC) analysis for ERP feature distributions as reported in Figures~\ref{fig:aucAFF}~and~\ref{fig:aucEFF}. The AUC scores, visualizing the EEG features separability for a subsequent classification, were higher ($AUC> 0.65$) for the efferent motion cases depicted in Figures~\ref{fig:aucEFF700}~and~\ref{fig:aucEFF150} for $ISI$ of $700$~ms and $150$~ms, respectively. The AUC analysis results were confirmed also with vmoBCI classification accuracies summarized in Tables~\ref{tab:aff}~and~\ref{tab:eff} for afferent and efferent vmoBCI online experiments using ten and five brainwave ERP averaging scenarios. The mean accuracies were non--significantly differing, as tested with pairwise rank--sum Wilcoxon method, among the tested afferent and efferent cases with various $ISI$ and averaging settings.
\begin{table}[h]
	\caption{Afferent visual motion onset BCI accuracy results (a chance level of $16.7\%$ due to six commands used)}\label{tab:aff}
	\begin{center}
	\begin{tabular}{c c c c c}
	\toprule
	\multirow{2}{*}{User number}	& \multicolumn{2}{c}{$ISI=700$~ms} & \multicolumn{2}{c}{$ISI=150$~ms} \\
	%\cline{2-5}
							& ~$10$ trials~ &  ~$5$ trials~	& ~$10$ trials~ &  ~$5$ trials~ \\
	\toprule
	$\#1$			 & $100.0\%$ & $~83.3\%$	& $100.0\%$ & $100.0\%$ \\
	$\#2$			 & $~16.7\%$ & $~50.0\%$	& $~66.7\%$ & $~83.3\%$ \\
	$\#3$			 & $100.0\%$ & $~83.3\%$	& $100.0\%$ & $~66.7\%$ \\
	$\#4$			 & $100.0\%$ & $100.0\%$	& $100.0\%$ & $~66.7\%$ \\
	$\#5$			 & $100.0\%$ & $100.0\%$	& $100.0\%$ & $~66.7\%$ \\
	\midrule
	Average accuracy & $83.3\pm37.2\%$ & $83.3\pm20.4\%$ & $93.3\pm14.9\%$ & $76.7\pm14.9\%$ \\
	\bottomrule
	\end{tabular}
	\end{center}
	\label{tab:endoACC}
\end{table}%
\begin{table}[h]
	\caption{Efferent visual motion onset BCI accuracy results (a chance level of $12.5\%$ due to eight commands used)}\label{tab:eff}
	\begin{center}
	\begin{tabular}{ c c c c c }
	\toprule
	\multirow{2}{*}{User number}	& \multicolumn{2}{c}{$ISI=700$~ms} & \multicolumn{2}{c}{$ISI=150$~ms} \\
	%\cline{2-5}
							& ~$10$ trials~ & ~$5$ trials~	& ~$10$ trials~ & ~$5$ trials~ \\
	\midrule
	$\#1$					& $100.0\%$ & $100.0\%$	& $100.0\%$ & $100.0\%$ \\
	$\#2$					& $100.0\%$ & $~75.0\%$	& $100.0\%$ & $~87.5\%$ \\
	$\#3$					& $~62.5\%$ & $~50.0\%$	& $~75.0\%$ & $~75.0\%$ \\
	$\#4$					& $100.0\%$ & $100.0\%$	& $100.0\%$ & $100.0\%$ \\
	$\#5$					& $~87.5\%$ & $~87.5\%$	& $~87.5\%$ & $~87.5\%$ \\
	\midrule
	Average accuracy	& $92.0\pm16.5\%$ & $82.5\pm20.9\%$ & $92.5\pm11.2\%$ & $90.0\pm10.5\%$ \\
	\bottomrule
	\end{tabular}
	\end{center}
	\label{tab:endoACC}
\end{table}%

%%%%%%%%%%%%%%%%%%%%%%%%%%%%%%%%%%%%%%%%%%
\section{Discussion}

The aim of this study was to test the ISI variability impact on the vmoBCI classification accuracy results in various brainwave averaging scenarios of the proposed visual motion onset paradigm. The online results obtained with the swLDA classifier did not show significant differences for the both afferent and efferent motion scenarios.

The results are very promising for future online applications with patients suffering from LIS allowing for speeding up the BCI stimuli presentation without significant classification degradation.

%%%%%%%%%%%%%%%%%%%%%%%%%%%%%%%%%%%%%%%%%%
\section{Materials and Methods}

The visual motion onset stimuli were delivered as colored moving objects on a computer display in moving towards the center (afferent) and off--center (efferent) trajectories as presented in Figures~\ref{fig:affVMOBCI}~and~\ref{fig:effVMOBCI}. The animations were programmed in Processing.org visual programming environment.  
There were six afferent and eight efferent pattens tested delivered in
random order in order to elicit P300 brainwave responses in an oddball style paradigm~\cite{bciBOOKwolpaw}.
The inter--stimulus intervals ($ISI$s) of $700$~ms and $150$~ms were chosen. The two different ISI setups were tested in order to evaluate a possible impact of the fast stimulation repetitions on user's BCI performance (the resulting BCI accuracy) in two different brainwave event related potential (ERP) averaging scenarios using ten or five responses. 
During the EEG experiments, the users were instructed to watch center of a computer screen and to minimize eye--movements as shown in Figures~\ref{fig:affVMOBCI}~and~\ref{fig:effVMOBCI}, as well as in online videos from the experiment\footnote{Online demo videos from the vmoBCI experiments: afferent -- \url{https://youtu.be/JbdazbFntek} and efferent -- \url{https://youtu.be/UaxIQdOctcQ} cases.}. The users responded mentally by confirming/counting only to the instructed visual motion patterns while ignoring the others. 
The EEG signals were captured with a portable EEG amplifier system g.USBamp (g.tec Medical Instruments, Austria). Eight active wet EEG electrodes were used to capture brainwaves with attentional modulation elucidated, within the event related potentials (ERPs),  the so--called ``aha-'' or P300-responses~\cite{bciBOOKwolpaw}. The EEG electrodes were attached to the head locations
\emph{CP3, CP4, POz, Pz, CPz, Cz, P3}, and \emph{P4} as in $10/10$
intentional system~. A reference
electrode was attached to a left earlobe and a ground electrode on the
forehead at \emph{FPz} position respectively. 

The details of the experimental procedures and the research targets of the tpBCI paradigm were explained in detail to the five human users, who agreed voluntarily to participate in the study.
The electroencephalogram (EEG) vmoBCI experiments were conducted in accordance with \emph{The World Medical Association Declaration of Helsinki - Ethical Principles for Medical Research Involving Human Subjects}. 
The experimental procedures were approved and designed in agreement with the ethical committee guidelines of the Faculty of Engineering, Information and Systems at University of Tsukuba, Tsukuba, Japan (experimental permission no.~$2013R7$). 
The average age of the users was of $26.8\pm10.7$ years.

The EEG signals were recorded and preprocessed online by an in--house extended  BCI2000 application~\cite{bci2000book,yoshihiroANDtomekAPSIPA2013} and segmented (``epoched'') as features drawn from ERP intervals of $0 \sim 700$~ms. 
The sampling rate was set to $512$~Hz, the high pass filter at $0.1$~Hz, and
the low pass filter at $40$~Hz. The ISI were of $700$~ms or $150$~ms in two different experimental runs.
Each user performed two sessions of selecting six or eight patterns (a spelling of a sequence of digits associated with each tactile pressure pattern) in afferent and efferent visual motion settings, respectively.
We performed offline analysis of the collected online EEG datasets in order to test a possible influence of the two ISI settings on the vmoBCI accuracy (compare ERP results in Figures~\ref{fig:erpAFF700},~\ref{fig:erpAFF150},~\ref{fig:erpEFF700},~and~\ref{fig:erpEFF150} depicting an impact of faster ISI on the ERP shapes). 
The swLDA~\cite{krusienski2006} classifier was applied next, with features drawn from the $0\sim700$~ms ERP intervals, with removal of the least significant input features, having $p > 0.15$, and with the final discriminant function restricted to contain a maximum of $60$ features.

%%%%%%%%%%%%%%%%%%%%%%%%%%%%%%%%%%%%%%%%%%
\section{Conclusions}

The approach presented shall help, if not to reach the goal, to get closer to our objective of the more user friendly visual BCI design. Thus, we can expect that patients suffering from LIS, as well as healthy users, will be able to use the more perceptually friendly BCI interfaces, according to their intact sensory modalities. We expect that the proposed visual motion onset paradigm shall offer a  more efficient and comfortable stimulus--driven BCI alternative.

Although the statistical analysis did not yield significant differences between the tested afferent and efferent motion cases, the users preferred the efferent stimuli resulting with better speeds as shown in the example with single ERP classification documented in an online video\footnote{Online demo video showing the efferent vmoBCI case in the single ERP averaging case -- \url{https://youtu.be/UaxIQdOctcQ}.}.

Still there remains a long path to go before providing a natural and comfortable visual motion onset BCI end--user solution, yet our research has progressed toward this goal as presented in this paper.

%%%%%%%%%%%%%%%%%%%%%%%%%%%%%%%%%%%%%%%%%%
\vspace{6pt} 

%%%%%%%%%%%%%%%%%%%%%%%%%%%%%%%%%%%%%%%%%%
%% optional
%\supplementary{The following are available online at www.mdpi.com/link, Figure S1: title, Table S1: title, Video S1: title.}

%%%%%%%%%%%%%%%%%%%%%%%%%%%%%%%%%%%%%%%%%%
%\acknowledgments{All sources of funding of the study should be disclosed. Please clearly indicate grants that you have received in support of your research work. Clearly state if you received funds for covering the costs to publish in open access.}

%%%%%%%%%%%%%%%%%%%%%%%%%%%%%%%%%%%%%%%%%%
\authorcontributions{TMR conceived the visual motion onset BCI paradigm; JPJ, CT and TMR designed and performed the experiments; TMR analyzed the data; TMR wrote the paper.}

%%%%%%%%%%%%%%%%%%%%%%%%%%%%%%%%%%%%%%%%%%
\conflictofinterests{The authors declare no conflict of interest.} 

%%%%%%%%%%%%%%%%%%%%%%%%%%%%%%%%%%%%%%%%%%
%% optional
\abbreviations{The following abbreviations are used in this manuscript:\\

\noindent BCI: Brain--computer interface\\
BMI: Brain--machine interface\\
ERP: Event--related potential\\
ISI: Inter--stimulus--interval\\
swLDA: Stepwise linear discriminant analysis classification\\
vmoBCI: Visual motion onset stimulus--based brain--computer interface}

%%%%%%%%%%%%%%%%%%%%%%%%%%%%%%%%%%%%%%%%%%
%% optional
%\appendix
%\section{}
%The appendix is an optional section that can contain details and data supplemental to the main text. For example, explanations of experimental details that would disrupt the flow of the main text, but nonetheless remain crucial to understanding and reproducing the research shown; figures of replicates for experiments of which representative data is shown in the main text can be added here if brief, or as Supplementary data. Mathemtaical proofs of results not central to the paper can be added as an appendix.
%
%\section{}
%All appendix sections must be cited in the main text. In the appendixes, Figures, Tables, etc. should be labeled starting with `A', e.g., Figure A1, Figure A2, etc. 

%%%%%%%%%%%%%%%%%%%%%%%%%%%%%%%%%%%%%%%%%%
%\bibliographystyle{mdpi}
%\bibliography{tomekONE,tomekCV}

%%%%%%%%%%%%%%%%%%%%%%%%%%%%%%%%%%%%%%%%%%
%% optional
%\sampleavailability{Samples of the compounds ...... are available from the authors.}

%%%%%%%%%%%%%%%%%%%%%%%%%%%%%%%%%%%%%%%%%%
\end{document}